\documentclass{article}
\usepackage{spconf,amsmath,graphicx,hyperref}
\usepackage{tabularx}
\usepackage{array}
\usepackage{lipsum}
\usepackage{comment}
\usepackage{multirow}
\usepackage{booktabs}
\title{DO BIOACOUSTIC FOUNDATION MODELS HEAR THROUGH THE NOISE? \newline SNR AND GENERALIZATION IN UNDERWATER DETECTION}
\name{Lucie JEAN-LABADYE$^{1,2,3}$, Nicolas FARRUGIA$^{3}$, Axel MARMORET$^{3}$, Dorian CAZAU$^{2}$, Olivier ADAM$^{1}$}

\address{$^{1}$Sorbonne Université, Institut $\partial$'Alembert, UMR CNRS 7190, F-75005, Paris, France\\
         $^{2}$ENSTA, Lab-STICC, UMR CNRS 6285, F-29806, Brest, France \\
         $^{3}$IMT Atlantique, Lab-STICC, UMR CNRS 6285, F-29238 Brest, France}
\begin{document}
%
\maketitle
\begin{abstract}
Bioacoustic foundation models are claimed to generalize across taxa and recording conditions, but are trained mainly on terrestrial archives of focal, high-quality recordings and evaluated on benchmarks emphasizing architecture over data conditions. We evaluate four such models as frozen encoders on an underwater detection task, using linear probing on continuously recorded data from the Kerguelen Plateau. Four datasets of identical size were built to differ only in the signal-to-noise ratio (SNR) of their positive segments, and probes were trained and evaluated across all combinations. Evaluation SNR accounts for more variation in average precision (AP) than model choice, and performance peaks when training and evaluation share a noise regime --- the configuration least representative of a continuous deployment. Training on mixed SNR redistributes performance toward low-SNR conditions without improving it overall, and AP remains low throughout. In light of these results, we call for evaluation standards grounded in realistic noise and deployment conditions rather than curated, high-quality recordings.
\end{abstract}

\section{Introduction \& prior work}
\label{sec:intro}
Passive acoustic monitoring (PAM) is a remote observation method well suited to ecosystems where visual surveys are difficult and where the species of interest are vocally active. It is therefore especially valuable for studying cetaceans, which rely on sound for all their vital activities. Their study is of growing importance as whale populations face sustained pressure from human activities at sea~\cite{cheesemanBellwethersChangePopulation}. Assessing such pressures requires monitoring over years, which PAM makes feasible through continuous, long-term campaigns. The resulting data volumes, however, make purely manual analysis intractable. The process is indeed time- and attention-consuming and requires substantial expertise to discriminate between sound sources.
Because it relies on expert judgment, the process is difficult to reproduce, and annotation quality varies between analysts~\cite{dubusBetterQuantifyingInterannotator2023}. These constraints have motivated the development of automatic methods.

Early automation of detection (\textit{is a signal of interest present?}), on which this study focuses, and classification (\textit{what is its source?}) relied on advanced signal processing methods such as spectrogram correlation~\cite{mellingerRecognizingTransientLowfrequency2000}. More recently, deep learning has substantially improved PAM data analysis~\cite{stowellComputationalBioacousticsDeep2022}, with automatic detectors and classifiers often outperforming traditional approaches~\cite{shiuDeepNeuralNetworks2020}. However, such models are often developed by practitioners for a specific geographic area, deployment period, or ecological question, and are neither designed nor evaluated for transfer beyond that setting. 
In parallel, following broader developments in deep learning, bioacoustic foundation models have begun to emerge~\cite{schwingerFoundationModelsBioacoustics2026}. Trained on large and taxonomically diverse datasets, these models are designed to address a range of tasks with little or no retraining, and to be inexpensive to specialize when adaptation is required. Their promise for PAM is therefore considerable: detectors and classifiers that generalize across geographic areas and recording conditions without site-specific training. 

Nonetheless, two obstacles seem to limit the adoption of foundation models for marine detection tasks. First, these models are predominantly trained on terrestrial data, where large multi-species archives such as Xeno-Canto~\cite{vellingaXenocantoCollectionIts2015a} are more readily compiled. Such archives consist largely of focal recordings, which differ markedly from continuous marine recordings. Available underwater datasets such as Watkins~\cite{sayighWatkinsMarineMammal2016} are, by contrast, often dated, distributed as focal clipped excerpts~\cite{burnsPerch20Transfers} --- suitable for classification but ill-suited to detection --- or only partially annotated (\textit{e.g.} restricted to a single species, vocalization, or family of vocalizations). Benchmarks such as BEANS~\cite{hagiwaraBEANSBenchmarkAnimal2023a} have considerably advanced the development and fair comparison of deep learning models, but do not make operability an explicit criterion. Second, benchmarks often emphasize models and their ability to generalize across biomes, with comparatively little attention paid to how data are partitioned between training, validation, and test sets, to the statistical differences between those partitions, or to data quality under operational conditions. As a result, reported performance does not necessarily reflect the conditions practitioners face.

This gap has consequences on both sides. Foundation models remain underused by the community that would benefit from them most, and the biases they carry remain uncharted. Unless these sources of bias and uncertainty are accounted for, trends in estimated call densities cannot be reliably attributed to biological drivers, such as changes in animal density or call production rates~\cite{millerCommonGroundEfficient}.
One data property is particularly consequential in this respect: the signal-to-noise ratio (SNR). By quantifying a signal's level relative to the surrounding background noise, SNR directly determines the difficulty of the detection task. It is also used by biologists and ecologists to model detection probability and, in turn, to estimate population density from acoustic data~\cite{decastroCountingCallsEstimating2024}.

Recent work~\cite{MICHAUD2025114475} has examined the impact of SNR on the performance of automatic detectors, showing that shifting the SNR distribution of the training set alters a model's ability to generalize. That study, however, used terrestrial recordings and task-specific models, whereas foundation models are explicitly claimed to offer superior generalization. To our knowledge, that claim has not been tested under the noise conditions of continuous underwater recordings.

This work then makes the following contributions. First, we use linear probing to evaluate the representations produced by four bioacoustic foundation models used as frozen encoders. Second, we conduct this evaluation on continuously recorded PAM data 
reflecting the conditions and class balance of operational deployments. Third, we assess generalization across controlled SNR strata, isolating noise level as a variability factor and quantifying its effect on detection performance. We hypothesize, first, that evaluations that do not control for data properties such as SNR give an optimistic picture of model ability; and second, that training on a mixture of SNR conditions generalizes better to realistic noise regimes than any single condition.

\begin{figure*}[t]
  \centering
  \includegraphics[width=0.8\textwidth]{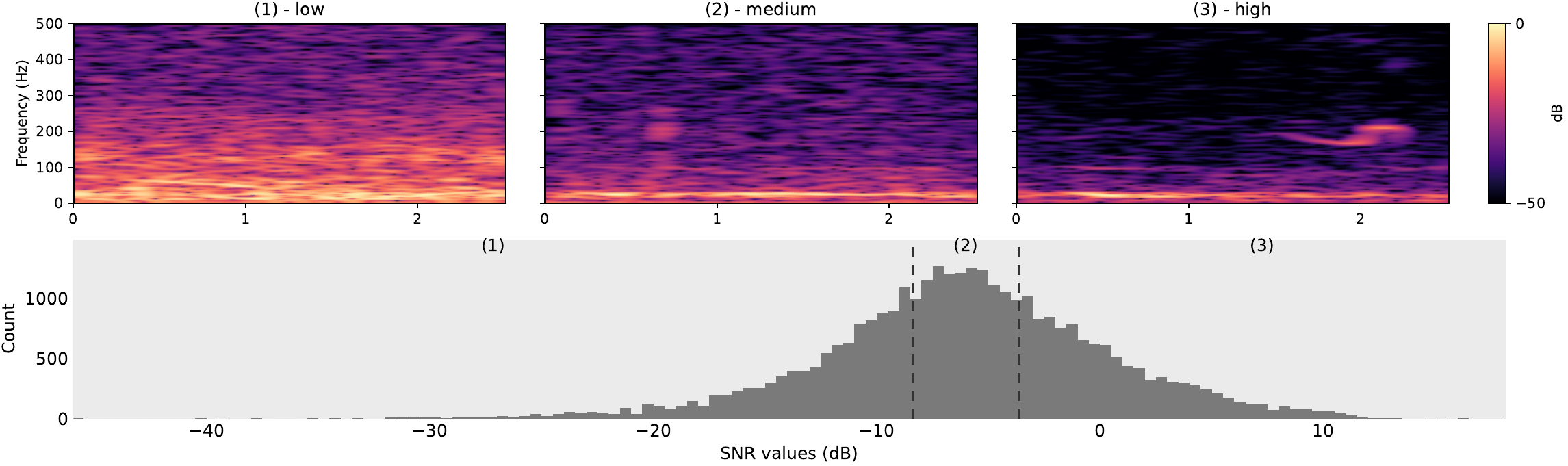}
  \vspace{-3mm}
  \caption{SNR distribution across the whole dataset, with an example spectrogram for each level}
  \label{fig:hist_spectro}
\end{figure*}

\section{MATERIAL \& METHODS}
\label{sec:material and methods}
\subsection{Data collection}
We used an open-access dataset, collected and annotated by Miller et al.~\cite{millerOpenAccessDataset2021} as part of the Acoustic Trends Project, which investigates the populations and dynamics of Antarctic blue and fin whales (\textit{Balaenoptera musculus intermedia} and \textit{Balaenoptera physalus}) in the Southern Ocean through acoustics. 
Recordings were acquired by a moored recorder deployed off the Kerguelen Plateau in 2014 and 2015. The hydrophone recorded continuously throughout each year, at a sampling rate of 12~kHz with 16-bits resolution. To limit the annotation workload, the authors uniformly subsampled 200 one-hour files per year and annotated those. Their annotation protocol targeted seven call types,\footnote{Blue whale: BmA, BmB, and BmZ; fin whale: Bp20 and Bp20plus; both species: BmDs and BpDs.} annotated as time--frequency bounding boxes on spectrograms. We pooled them into a single presence/absence label per segment, framing the problem as weakly labeled detection.

\subsection{Data preprocessing}
To match the input format expected by the pre-trained encoders, each one-hour recording was divided into 5~s segments with 50\% overlap, yielding 1440 segments per file. A segment was considered positive if it contained at least 50\% of the duration of an overlapping sound event. Note that the resulting dataset is strongly imbalanced, with 96\% of the segments labeled negative.

Following Miller et al.~\cite{millerOpenAccessDataset2021}, the SNR of each annotated event was computed as in~\cite[Eq. 6.26]{IntroductionUnderwaterAcoustics}, which defines the output SNR of a quadratic receiver in natural values. Equation~\ref{eq:snr} expresses this result in decibels:
\begin{equation}
\label{eq:snr}
\mathrm{SNR}_{\mathrm{dB}} = 10 \log_{10}
    \left( \frac{\langle z_{s+n}(t) \rangle - \langle z_{n}(t) \rangle}
                {\sqrt{\sigma^{2}_{n}}} \right)
\end{equation}
where $z_{s+n}$ is the power within the annotation box, computed from the spectrogram; $z_{n}$ is the power of a noise-only window of equal bandwidth, taken 5~s before the annotation box and separated from it by a 1~s buffer to accommodate variability in the annotators' box edges; and $\sigma_n^2$ is the variance of the noise power over that window, homogeneous to its power under white-noise assumption. Finally, $\langle \cdot \rangle$ denotes the temporal mean. The selection of this SNR definition was informed by preliminary studies that fall outside the scope of this work.

When a segment overlapped several events, it inherited the SNR of the event with which it overlapped the most. Positive segments were then divided into three equal-size SNR levels, each containing one third of the positive segments. The resulting boundaries are $-8.66$~dB between low and medium levels and $-4.01$~dB between medium and high levels. Fig.~\ref{fig:hist_spectro} shows their distribution.

\subsection{Pre-trained models}
The foundation models used in this study were published in 2025--2026. Two of them come from the Earth Species Project (ESP) and their AVEX library~\cite{hagiwaraAVESAnimalVocalization2023}: an EfficientNet B0 (esp-effnet) and a Transformer based on Audio Transformer (esp-eat). Both were trained on a mixture of bioacoustic datasets (Xeno-Canto~\cite{vellingaXenocantoCollectionIts2015a}, iNaturalist~\cite{hornINaturalistSpeciesClassification2018}, Watkins~\cite{sayighWatkinsMarineMammal2016}, and Animal Sound Archive~\cite{frommoltAnimalSoundArchive2006}) and on AudioSet~\cite{gemmekeAudioSetOntology2017}, a general-purpose corpus spanning speech, music, and environmental sounds. The two other models were published by Google: Perch2.0 (perch)~\cite{merrienboerPerch20Bittern2025} and multi-species-whales (msw)~\cite{harrellWhistlesSongsBoings2024}. Both are also EfficientNet-based (B0 for msw, B3 for perch), but were trained on different data mixtures. Perch was trained on Xeno-Canto, iNaturalist and Animal Sound Archive, and on the general-purpose FSD50k~\cite{fonsecaFSD50KOpenDataset2022}, whereas msw was mostly trained on underwater acoustic data compiled by NOAA~\cite{pifscPIPAN10kHz2021}. 

\begin{table}[t] 
\centering 
\small 
\begin{tabular}{@{}lrrr@{}} 
\toprule 
& Train & Val. & Test \\ 
\midrule 
Positive segments & 6996 & 1749 & 2187 \\ 
Negative segments & 6996 & 1749 & 36187 \\ 
Control, per stratum & 2332 & 583 & 729 \\ 
\bottomrule 
\end{tabular} 
\caption{Data splits. All four conditions have the same size and differ only in the SNR of their positives.} \label{tab:splits} 
\end{table}

\subsection{Experimental set-up}
We built four datasets, identical in size and stratified across labels before binarization, differing only in the SNR of their positive segments. Three are SNR-specific and contain positives drawn only from a single level (low, medium or high). The fourth serves as a control and draws its positives in equal proportion from the three levels, to ensure no impact of data volume. Negative segments were drawn from a common pool.

Each dataset was then split into training, validation, and test sets. To mitigate class imbalance, the training and validation sets were balanced to a 1:1 positive-to-negative ratio by subsampling negatives, whereas the test sets retained the original class distribution to reflect realistic deployment conditions. The resulting splits are detailed in Table~\ref{tab:splits}.

To match the sampling rate expected by each model, the 5~s segments were resampled from 12 kHz to 16 kHz for the ESP models, to 24 kHz for msw, and to 32 kHz for perch. Segments were then encoded, and the resulting embeddings were evaluated through linear probing. Since the aim of this study is to assess whether SNR affects the generalization ability of the embeddings themselves, a more expressive downstream model would have confounded representation quality with the capacity of the classifier trained on top of it.

A linear probe was trained on each of the four datasets, using 5-fold cross-validation to characterize variability across folds, and evaluated on the test sets of all four conditions. This yields the full matrix of training and evaluation SNR combinations reported in Fig.~\ref{fig:snr}. Performance is reported as average precision (AP), which summarizes the precision–recall curve without committing to a single operating threshold~\cite{hildebrandPerformanceMetricsMarine2022}.

\section{RESULTS \& DISCUSSION}
\label{sec:results}
Each factor (model, training SNR, and test SNR) was tested as a separate one-way repeated-measures ANOVA on marginal means, with fold as the repeated-measures unit and AP as the dependent variable. As a mixture, the control condition was excluded from the SNR marginals and tested through planned contrasts (t-tests) against the mean of the three pure levels.

\subsection{Model comparison}
\begin{figure}[t]
  \centering
  \includegraphics[width=0.9\linewidth]{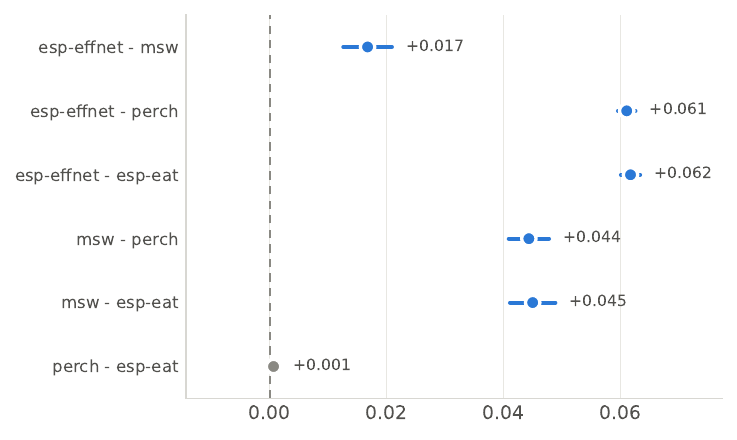}
  \vspace{-3mm}
  \caption{Pairwise AP differences between models}
  \label{fig:models}
\end{figure}

Fig.~\ref{fig:models} reports pairwise differences between the four models, averaged across SNR conditions. The main effect of model on AP is significant $\left(F(3, 12) = 1831, \text{p}_{\text{holm}} < 10^{-5}\right)$. In particular, esp-effnet outperforms the three others by $+0.017$, $+0.061$, and $+0.062$ over msw, perch, and esp-eat, respectively $ \left(\text{all}~\text{p}_{\text{holm}} < 10^{-3}\right)$. We thus retain esp-effnet for the remaining analysis.

Yet differences between models remain modest compared to the impact of the SNR levels: across the marginal means, test sets span 0.139 AP, more than twice the spread observed between models (0.062). The following therefore focuses on how SNR shapes detection performance.

\subsection{Effects of the SNR}

\begin{figure}[t]
  \centering
  \includegraphics[width=0.7\linewidth]{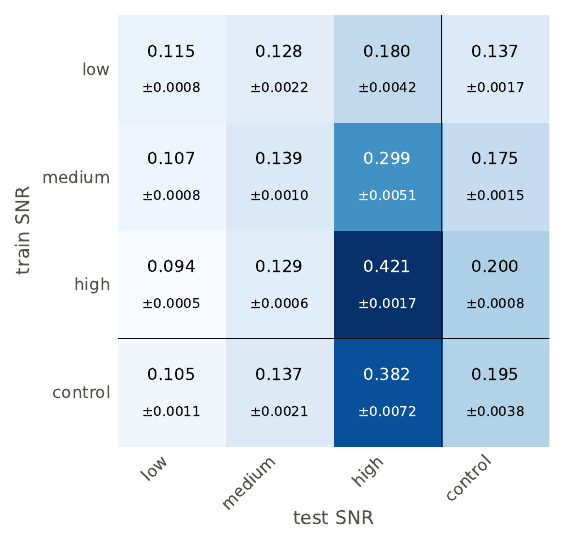}
  \vspace{-3mm}
  \caption{AP of esp-effnet across training and evaluation SNR}
  \label{fig:snr}
\end{figure}

Fig.~\ref{fig:snr} reports esp-effnet's AP for every combination of training and evaluation strata. Three patterns emerge that are consistent across models.

First, AP increases monotonically with test SNR ($F(2, 8) = 3.9 \times 10^4$, partial $\eta^2 = 0.999$), irrespective of training stratum: a probe trained only on low-SNR examples still improves from 0.115 on low-SNR data to 0.180 on high-SNR data.

Second, this relationship is asymmetric. Exposure to high-SNR examples during training does not transfer downwards: the high-trained probe reaches 0.094 on low-SNR data, below both the medium-trained (0.107) and the low-trained (0.115) probes. Training on easy examples therefore costs performance where detection is hardest.

Third, performance is highest when the training and evaluation strata match. Averaged over models, matched SNR levels exceed mismatched ones by $+0.049$ AP beyond what the main effects alone predict, and performance decays further as the gap widens ($+0.047$ AP for a one-step versus a two-step mismatch, \textit{e.g.} low--medium versus low--high). Each of the low, medium, and high test columns is maximized on the diagonal, an oracle condition that assumes the noise regime of the deployment is known in advance. The exception is the control column, where the high-trained probe (0.200) marginally exceeds the control-trained one (0.195).

Exposure to a range of SNR levels during training nonetheless improves generalization overall, with a mean gain of $+0.0221$ across models and evaluation strata $(t(4) = 33.78, \text{p}_{\text{holm}} < 10^{-5})$. For esp-effnet, the benefit is concentrated where it matters: the control-trained probe outperforms the high-trained one on low- (0.105 vs. 0.094) and medium-SNR (0.137 vs. 0.129) evaluation data, while giving up ground on high-SNR data (0.382 vs. 0.421).

\subsection{Discussion}

Training and evaluating on high-SNR data yields an AP of 0.421, more than four times that obtained when the same probe is evaluated on low-SNR data (0.094). Yet high-SNR training and evaluation is the configuration most likely to be reported in benchmark settings and the least representative of deployment. Recordings from a continuously deployed mooring are dominated by distant, low-SNR calls, and the corresponding cells of Fig.~\ref{fig:snr} fall between 0.094 and 0.115. Absolute values are low throughout, and we take this as the central result: on continuous, long-term marine PAM data, current bioacoustic foundation models do not deliver the out-of-the-box generalization their framing suggests. Moreover, given that the effect of the test set exceeds that of the model, how a model is evaluated seems to matter more than which model is evaluated. This argues for future work in which test sets are characterized as carefully as architectures are compared.

Another factor plausibly contributes: low-SNR calls are harder to annotate, so the ground truth is likely least complete where performance is lowest. A preliminary visual inspection of high-confidence estimated detections (using esp-effnet and perch) supports this: several spectrograms display patterns consistent with calls that remain unannotated. Part of the gap may thus reflect annotation completeness rather than model capability. This motivates active learning pipelines: rather than replacing expert analysts, such models could support PAM post-processing by directing annotation effort toward the most informative examples, helping complete the ground truth where it matters most.

\section{CONCLUSION}
\label{sec:conclusion}

By isolating SNR as a single controlled factor, this study shows that the noise conditions of the evaluation data account for more variation in detection performance than the choice of foundation model. Performance is also highest when training and evaluation share a noise regime --- a condition rarely met in practice, since the noise conditions of a long-term deployment are neither known in advance nor stable. 

Multiple directions follow. Extending this protocol to other sources of variability --- geographic area, time period, recording hardware --- would establish whether the observed patterns generalize beyond SNR level. Detections on unannotated segments also suggest an active learning pipeline in which model confidence is used to direct annotation effort toward the faintest calls.

These results argue for evaluating bioacoustic foundation models on the data conditions practitioners actually encounter. Bridging the two communities matters, as acoustic monitoring directly feeds conservation policies. Given the pressure cetaceans are under, characterizing foundation models' biases is an immediate practical requirement, not a refinement for later work.

\vfill\pagebreak

\bibliographystyle{IEEEbib}
\bibliography{refs}

@article{burnsPerch20Transfers,
  title = {Perch 2.0 Transfers `Whale' to Underwater Tasks},
  author = {Burns, Andrea and Harrell, Lauren and {van Merri{\"e}nboer}, Bart and others},
  year = 2025,
  month = dec,
  howpublished = {arXiv:2512.03219},
}

@article{decastroCountingCallsEstimating2024,
  title = {Beyond Counting Calls},
  author = {{de Castro}, Franciele R. and Harris, Danielle V. and Buchan, Susannah J. and Balcazar, Naysa and Miller, Brian S.},
  year = 2024,
  month = sep,
  journal = {Frontiers in Marine Science},
  volume = {11},
  publisher = {Frontiers},
  issn = {2296-7745},
  doi = {10.3389/fmars.2024.1406678},
  urldate = {2025-11-28}
}

@inproceedings{dubusBetterQuantifyingInterannotator2023,
  title = {Better Quantifying Inter-Annotator Variability: {{A}} Step towards Citizen Science in Underwater Passive Acoustics},
  shorttitle = {Better Quantifying Inter-Annotator Variability},
  booktitle = {{{OCEANS}} 2023 - {{Limerick}}},
  author = {Dubus, Gabriel and Torterotot, Ma{\"e}lle and Duc, Paul Nguyen Hong and Beesau, Julie and others},
  year = 2023,
  month = jun,
  pages = {1--8},
  publisher = {IEEE},
  doi = {10.1109/OCEANSLimerick52467.2023.10244502},
  urldate = {2026-09-15}
}

@misc{fonsecaFSD50KOpenDataset2022,
  title = {{{FSD50K}}: {{An Open Dataset}} of {{Human-Labeled Sound Events}}},
  shorttitle = {{{FSD50K}}},
  author = {Fonseca, Eduardo and Favory, Xavier and Pons, Jordi and others},
  year = 2022,
  month = apr,
  howpublished = {arXiv:2010.00475},
}

@book{frommoltAnimalSoundArchive2006,
  title = {The Animal Sound Archive at the {{Humboldt-University}} of {{Berlin}}},
  shorttitle = {The Animal Sound Archive at the {{Humboldt-University}} of {{Berlin}}},
  author = {Frommolt, Karl-Heinz and Bardeli, Rolf and Kurth, Frank and Clausen, Michael},
  year = 2006,
  publisher = {Slovenska akademija znanosti in umetnosti},
  urldate = {2026-09-15}
}

@inproceedings{gemmekeAudioSetOntology2017,
  title = {Audio {{Set}}: {{An}} Ontology and Human-Labeled Dataset for Audio Events},
  shorttitle = {Audio {{Set}}},
  booktitle = {2017 {{IEEE International Conference}} on {{Acoustics}}, {{Speech}} and {{Signal Processing}} ({{ICASSP}})},
  author = {Gemmeke, Jort F. and Ellis, Daniel P. W. and Freedman, Dylan and others},
  year = 2017,
  month = mar,
  pages = {776--780},
  publisher = {IEEE},
  address = {New Orleans, LA},
  doi = {10.1109/ICASSP.2017.7952261},
  urldate = {2026-09-15}
}

@inproceedings{hagiwaraAVESAnimalVocalization2023,
  title = {{{AVES}}: {{Animal Vocalization Encoder Based}} on {{Self-Supervision}}},
  shorttitle = {{{AVES}}},
  booktitle = {{{ICASSP}} 2023},
  author = {Hagiwara, Masato},
  year = 2023,
  month = jun,
  pages = {1--5},
  issn = {2379-190X},
  doi = {10.1109/ICASSP49357.2023.10095642},
  urldate = {2026-04-24}
}

@inproceedings{hagiwaraBEANSBenchmarkAnimal2023a,
  title = {{{BEANS}}: {{The Benchmark}} of {{Animal Sounds}}},
  author = {Hagiwara, Masato and Hoffman, Benjamin and Liu, Jen-Yu and others},
  shorttitle = {{{BEANS}}},
  booktitle = {{{ICASSP}} 2023},
  year = 2023,
  month = jun,
  pages = {1--5},
  publisher = {IEEE},
  address = {Rhodes Island, Greece},
  issn = {2379-190X},
  doi = {10.1109/ICASSP49357.2023.10096686},
  urldate = {2026-05-28}
}

@misc{hornINaturalistSpeciesClassification2018,
  title = {The {{iNaturalist Species Classification}} and {{Detection Dataset}}},
  author = {Horn, Grant Van and Aodha, Oisin Mac and Song, Yang and others},
  year = 2018,
  month = apr,
  howpublished = {arXiv:1707.06642},
  eprint = {1707.06642},
  primaryclass = {cs.CV},
  publisher = {arXiv},
  doi = {10.48550/arXiv.1707.06642},
  urldate = {2026-09-15},
}

@book{IntroductionUnderwaterAcoustics,
  title = {An {{Introduction}} to {{Underwater Acoustics}}},
  author = {Lurton, Xavier},
  year = 2010, 
  urldate = {2026-09-15},
}

@article{mellingerRecognizingTransientLowfrequency2000,
  title = {Recognizing Transient Low-Frequency Whale Sounds by Spectrogram Correlation},
  author = {Mellinger, David K. and Clark, Christopher W.},
  year = 2000,
  month = jun,
  journal = {J. Acoust. Soc. Am.},
  volume = {107},
  number = {6},
  pages = {3518--3529},
  issn = {0001-4966},
  doi = {10.1121/1.429434},
  urldate = {2026-09-15},
}

@misc{merrienboerPerch20Bittern2025,
  title = {Perch 2.0: {{The Bittern Lesson}} for {{Bioacoustics}}},
  shorttitle = {Perch 2.0},
  author = {van Merri{\"e}nboer, Bart and Dumoulin, Vincent and Hamer, Jenny others},
  year = 2025,
  month = aug,
  howpublished = {arXiv:2508.04665},
  eprint = {2508.04665},
  primaryclass = {cs},
  publisher = {arXiv},
  doi = {10.48550/arXiv.2508.04665},
  urldate = {2025-08-12}
}

@article{millerOpenAccessDataset2021,
  title = {An Open Access Dataset for Developing Automated Detectors of {{Antarctic}} Baleen Whale Sounds},
  author = {Miller, Brian S. and {The IWC-SORP/SOOS Acoustic Trends Working Group} and others},
  year = 2021,
  month = jan,
  journal = {Scientific Reports},
  volume = {11},
  number = {1},
  pages = {806},
  issn = {2045-2322},
  doi = {10.1038/s41598-020-78995-8},
  urldate = {2024-02-03}
}

@article{sayighWatkinsMarineMammal2016,
  title = {The {{Watkins Marine Mammal Sound Database}}: {{An}} Online, Freely Accessible Resource},
  shorttitle = {The {{Watkins Marine Mammal Sound Database}}},
  author = {Sayigh, Laela and Daher, Mary Ann and Allen, Julie and others},
  year = 2016,
  month = jul,
  journal = {Proceedings of Meetings on Acoustics},
  volume = {27},
  number = {1},
  pages = {040013},
  issn = {1939-800X},
  doi = {10.1121/2.0000358},
  urldate = {2026-09-15}
}

@article{schwingerFoundationModelsBioacoustics2026,
  title = {Foundation Models for Bioacoustics -- {{A}} Comparative Review},
  author = {Schwinger, Raphael and Vali Zadeh, Paria and Rauch, Lukas and others},
  year = 2026,
  month = jun,
  journal = {Ecological Informatics},
  volume = {96},
  pages = {103765},
  issn = {1574-9541},
  doi = {10.1016/j.ecoinf.2026.103765},
  urldate = {2026-09-15}
}

@article{shiuDeepNeuralNetworks2020,
  title = {Deep Neural Networks for Automated Detection of Marine Mammal Species},
  author = {Shiu, Yu and Palmer, K. J. and Roch, Marie A. and others},
  year = 2020,
  month = jan,
  journal = {Scientific Reports},
  volume = {10},
  number = {1},
  pages = {607},
  issn = {2045-2322},
  doi = {10.1038/s41598-020-57549-y},
  urldate = {2024-07-01},
}

@article{stowellComputationalBioacousticsDeep2022,
  title = {Computational Bioacoustics with Deep Learning: A Review and Roadmap},
  shorttitle = {Computational Bioacoustics with Deep Learning},
  author = {Stowell, Dan},
  year = 2022,
  month = mar,
  journal = {PeerJ},
  volume = {10},
  pages = {e13152},
  issn = {2167-8359},
  doi = {10.7717/peerj.13152},
  urldate = {2024-02-03}
}

@article{vellingaXenocantoCollectionIts2015a,
  title = {The {{Xeno-canto Collection}} and Its {{Relation}} to {{Sound Recognition}} and {{Classification}}.},
  author = {Vellinga, Willem-Pier and Planqu{\'e}, Robert},
  year = 2015,
  journal = {CLEF (Working Notes)},
  volume = {1391},
  urldate = {2026-09-15}
}

@article{millerCommonGroundEfficient,
  title = {Common Ground},
  year = 2026,
  month = may,
  author = {Miller, Brian S. and Kelly, Nat and Aulich, Meghan G. and Balcazar, Naysa and others},
  journal = {Methods in Ecology and Evolution},
  issn = {2041-210X},
  doi = {10.1111/2041-210x.70331},
  urldate = {2026-06-01}
}

@article{cheesemanBellwethersChangePopulation,
  title = {Bellwethers of Change},
  shorttitle = {Bellwethers of Change},
  author = {Cheeseman, Ted and Barlow, Jay and Acebes, Jo Marie and others},
  journal = {Royal Society Open Science},
  volume = {11},
  number = {2},
  pages = {231462},
  issn = {2054-5703},
  doi = {10.1098/rsos.231462},
  urldate = {2026-09-15}
}

@article{hildebrandPerformanceMetricsMarine2022,
  title = {Performance Metrics for Marine Mammal Signal Detection and Classification},
  author = {Hildebrand, John A. and Frasier, Kaitlin E. and Helble, Tyler A. and Roch, Marie A.},
  year = 2022,
  month = jan,
  journal = {J. Acoust. Soc. Am.},
  volume = {151},
  number = {1},
  pages = {414--427},
  issn = {0001-4966, 1520-8524},
  doi = {10.1121/10.0009270},
  urldate = {2024-03-01}
}

@misc{pifscPIPAN10kHz2021,
  author       = {{NOAA Pacific Islands Fisheries Science Center}},
  title        = {Pacific Islands Passive Acoustic Network ({PIPAN}) 10\,{kHz} Data},
  year         = {2021},
  publisher    = {NOAA National Centers for Environmental Information},
  doi          = {10.25921/Z787-9Y54},
  note         = {Accessed: 2026-09-15}
}

@misc{harrellWhistlesSongsBoings2024,
  author       = {Harrell, Lauren},
  title        = {Whistles, Songs, Boings, and Biotwangs: Recognizing Whale Vocalizations with {AI}},
  year         = {2024},
  month        = sep,
  howpublished = {Google Research Blog, https://research.google/blog/whistles-songs-boings-and-biotwangs-recognizing-whale-vocalizations-with-ai/},
  note         = {Accessed: 2026-09-15}
}

@article{MICHAUD2025114475,
title = {Acoustic detection of a nocturnal bird with deep learning: the challenge of low signal-to-noise ratio},
journal = {Ecological Indicators},
volume = {181},
pages = {114475},
year = {2025},
month = dec,
issn = {1470-160X},
doi = {https://doi.org/10.1016/j.ecolind.2025.114475},
url = {https://www.sciencedirect.com/science/article/pii/S1470160X25014074},
author = {Félix Michaud and Jérôme Sueur and Frédéric Sèbe and others}
}

\end{document}